\documentclass[preprint]{revtex4}
\usepackage{graphicx}
\usepackage{amsmath}
\usepackage{amssymb}
\usepackage{citesort}
\usepackage{subfigure}

\begin{document}

\title{Preferred sizes and ordering in surface nanobubble populations}

\author{Bram M. Borkent$^{\ddag}$, Holger Sch\"onherr$^{\dag}$\footnote{Present address: Physical Chemistry, University of Siegen, Adolf-Reichwein-Str. 2, 57076 Siegen, Germany.}, G{\'e}rard~Le~Ca{\"e}r$^{\flat}$, Benjamin Dollet$^{\flat}$, and Detlef~Lohse$^{\ddag}$}

\affiliation{
$^{\ddag}$Physics of Fluids, Faculty of Science and Technology and J.M. Burgers Centre for Fluid Dynamics and MESA$^+$ Institute for Nanotechnology, University of Twente, P.O. Box 217, 7500 AE Enschede, The Netherlands.\\ 
$^{\dag}$ Materials Science and Technology of Polymers, Faculty of Science and Technology and MESA$^+$ Institute for Nanotechnology, University of Twente, P.O. Box 217, 7500 AE Enschede, The Netherlands.\\ 
$^{\flat}$ Institut de Physique de Rennes, UMR UR1-CNRS 6251, Universit{\'e} de Rennes I, Campus de Beaulieu, B{\^a}timent 11A, F-35042 Rennes Cedex, France.
}
%
%
%
%

\date{\today}

\begin{abstract}
Two types of homogeneous surface nanobubble populations, created by different means, are analyzed statistically on both their sizes and spatial positions. In the first type (created by droplet-deposition, case A) the bubble size $R$ is found to be distributed according to a generalized gamma law with a \emph{preferred radius} $R^*=20$\,nm. The radial distribution function shows a \emph{preferred spacing} at $\sim 5.5 R^*$. These characteristics do not show up in comparable Monte-Carlo simulations of random packings of hard disks with the same size distribution and the same density, suggesting a structuring effect in the nanobubble formation process. The nanobubble size distribution of the second population type (created by ethanol-water exchange, case B) is a mixture of two clearly separated distributions, hence, with two preferred radii. The local ordering is less significant, due to the looser packing of the nanobubbles.
\end{abstract}



\maketitle

The first atomic force microscopy (AFM) observations of spherical cap-like soft domains at the solid-liquid interface~\cite{lou00,ishida00,yakubov00,carambassis98,tyrrell01}, later termed "surface nanobubbles", identified two typical, yet poorly understood, nanobubble characteristics: long-term stability and huge nanoscopic contact angles (on the water side). Later experiments confirmed these puzzling features of surface nanobubbles, and focused on verifying their gaseous nature by correlating the nanobubble densities with the gas concentration in the liquid~\cite{zhang04,zhang06c,yang07}. Recently, the gas content of the bubbles was identified explicitly by infrared spectroscopy measurements in combination with AFM~\cite{zhang07,zhang08a}. Other studies investigated the effect of surface active solutes~\cite{zhang06a,yang07}, salts~\cite{zhang06a}, substrate morphology~\cite{yang08}, or electrolysis~\cite{zhang06b,yang09} on the appearance, stability, and shape of surface nanobubbles. While the number of experiments supporting the notion that the observed structures are indeed surface nanobubbles, is  increasing~\cite{steitz03,switkes04,simonsen04,agrawal05,borkent07b,kameda08a,kameda08b,nam08}, no consensus has been reached concerning the mechanism which stabilizes the bubbles (see~\cite{brenner08} and references therein). One of the hypotheses, recently put forward in~\cite{brenner08}, is based on a non-stationary equilibrium between a gas outflux (through the gas-liquid interface) and a gas influx (at the three-phase contact line), and predicts a preferred nanobubble radius as a function of gas concentration and contact angle.

 In this paper we want to test the prediction~\cite{brenner08} of a preferred radius $R^*$ and its dependence on the gas concentration. Our good bubble statistics allow us to extract statistical properties of the whole nanobubble population. The analysis shows not only a preferred radius, but also a preferred spacing between the bubbles, suggesting a structuring mechanism between individual bubbles.

As substrates small pieces diced from a Si(100) wafer are used, which are subsequently cleaned, coated with a monolayer of 1$H$,1$H$,2$H$,2$H$-perfluorodecyldimethylchlorosilane and analyzed following the procedure described previously~\cite{borkent07b}. The static macroscopic contact angle is typically around $92\,^\mathrm{o}$. The substrates are then mounted in an atomic force microscope (VEECO/Digital Instruments (DI) multimode) equipped with a NanoScope IIIa controller (DI, Santa Barbara, CA) and measured in tapping mode in water using a DI liquid cell and V-shaped Si$_3$N$_4$ cantilevers, with spring constants of $0.3-0.5\,\mathrm{N}/\mathrm{m}$ (Nanoprobes, DI). The amplitude setpoint was chosen as high as possible, typically $>90\%$.
The size of the nanobubbles is extracted from the raw AFM topography images by application of a height-threshold~\cite{yang09}, which yields the location and radius $R$ of each nanobubble. The results are corrected for the finite size of the tip (R$_{\mathrm{tip}}$=20\,nm), as done elsewhere~\cite{kameda08b}. We note that the tip correction does not affect the conclusions of this paper qualitatively.

\begin{figure}
\centering
    \includegraphics[width=8cm]{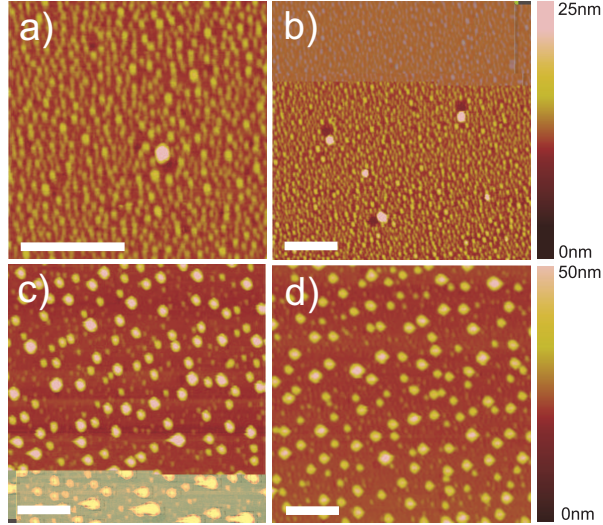}
    \caption{\label{Fig:cases} AFM topography images of the solid-liquid interface of the substrates. In a) and b) gas-equilibrated MilliQ-water was put on the substrate without explicit use of local oversaturation (case A1 and A2, resp.). In c) and d) the result is shown after a local and temporal oversaturation has been applied (case B1 and B2, resp.). Each scale bar corresponds to $1 \mu\mathrm{m}$.}
\end{figure}

The populations of surface nanobubbles are created in two different ways: in case A, a drop of gas-equilibrated Milli-Q water is put on the substrate, while in case B a finite, temporal local gas oversaturation (by flushing ethanol away with water~\cite{lou00,zhang06a,yang07,borkent07b}) is employed to explicitly stimulate nanobubble formation. In both cases two typical images are selected which were suitable for further statistical analysis (see Fig.~\ref{Fig:cases}).
In Case A (Fig.~\ref{Fig:cases}a and b) a dense coverage of relatively small and rather uniformly sized nanobubbles is observed. This observation is not evident as not all labs find the 'spontaneous' occurrence of nanobubbles (see for instance the remark in~\cite{zhang08a} and references therein). Only incidently, some larger nanobubbles are visible, which are present next to a bubble-free area. Presumably, smaller nanobubbles have merged to these larger objects. A mixed population of both small and large nanobubbles can be created when a forced local oversaturation is applied temporally~\cite{zhang06a}, as shown in Fig.~\ref{Fig:cases}c and d (cases B1 and B2, resp.). After the local gas-oversaturation, the bulk gas concentration is restored to its equilibrium value. In addition, the bubbles have been exposed to a single shockwave, as described in~\cite{borkent07b}. We noticed that the large nanobubbles did not vanish or shrink during the course of the experiment (i.e. within a few hours).

\begin{figure}
\centering
    \includegraphics[width=9cm]{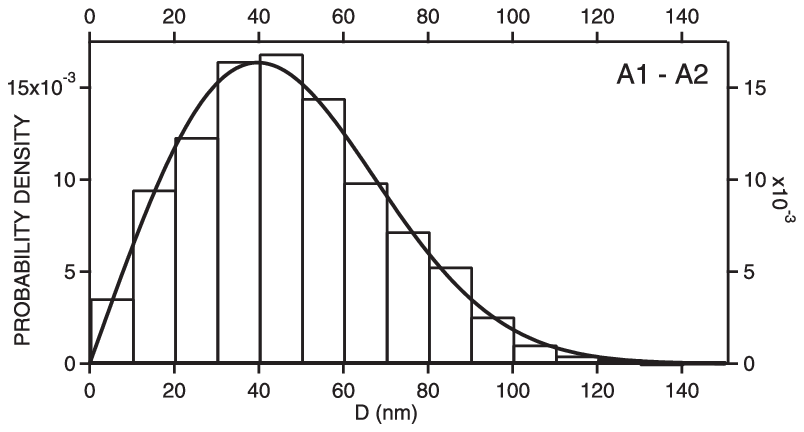}
    \includegraphics[width=9cm]{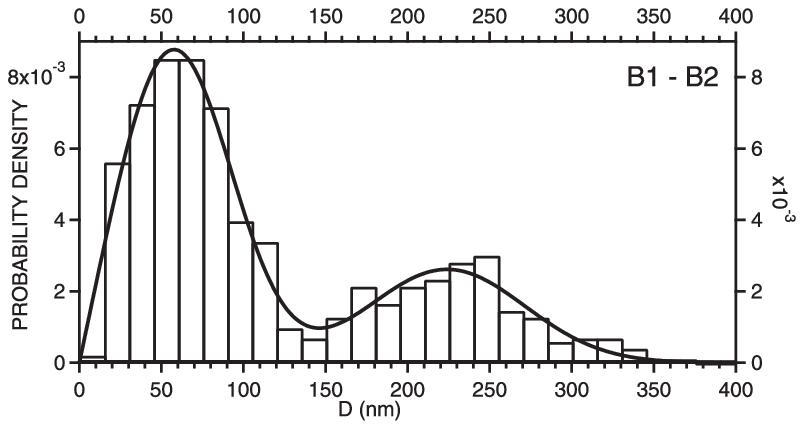}
    \caption{\label{Fig:pdf}Probability distribution of the nanobubble diameter $D$ in both case A (top) and case B (bottom). Each case is represented by two unique images (1 and 2, resp.), of which the total size distribution is shown. The bars depict experimental data, the lines show the best-fitted probability distribution.}
\end{figure}

The experimental probability size distributions $P(D)$ present in case A and B are shown in Fig.~\ref{Fig:pdf}. The bubble sizes clearly show a maximum at a particular diameter value, which we denote as the preferred diameter $D^*(=2R^*)$. In case B there are even two peaks, corresponding to two preferred radii. To obtain the value of $D^*$ the experimental size histograms were fitted with a generalized gamma distribution (GG)~\cite{stacy62} in case A and with a mixture of a GG and a Gaussian distribution in case B. In Case A the GG distribution which best fits the experimental results (Fig.~\ref{Fig:pdf}) is

\begin{equation}
P_A(D) = \frac{\theta_A \beta_A^2}{\Gamma(2/\theta_A)}D \exp \{-(\beta_A D)^{\theta_A}\},
\end{equation}

\noindent where $\Gamma (x)$ is Euler's gamma function, and $\beta_A$ and $\theta_A$ are shape parameters which are fitted, yielding $\beta_A=(1.73\pm0.07) \cdot 10^{-2}\,\mathrm{nm}^{-1}$ and $\theta_A=2.37\pm0.17$. As the value of the exponent of $D$ in front of the exponential was found to be very close to 1, it was fixed to 1. The maximum (or the mode) of $P_A(D)$ is formed at $D^*_A=1/\beta_A \theta_A^{1/\theta_A}=40\pm2$\,nm and the mean diameter $\langle D \rangle_A =47\pm2$\,nm. The standard deviation of the size distribution is $\sigma_A=23\pm2$\,nm.

In case B the total probability distribution could be fitted with a mixture of a GG distribution with the same form as that of $P_A(D)$ and of a Gaussian distribution:

\begin{equation}
P_B(D) = \alpha \frac{\theta_B \beta_B^2}{\Gamma(2/\theta_B)}D \exp \{-(\beta_B D)^{\theta_B}\} + \frac{1-\alpha}{\sigma_{B,2}\sqrt{2\pi}}\exp\left(-\frac{(D-D^*_{B,2})^2}{2\sigma_{B,2}^2}\right).
\end{equation}

\noindent As in case A, the exponent of $D$ was found to be close to 1 and fixed to that value. The fitted parameters are then $\alpha=0.69\pm0.04$, $\beta_B=(1.20\pm0.07) \cdot 10^{-2}\,\mathrm{nm}^{-1}$, $\theta_B=2.8\pm0.3$, $D_{B,2}^*=\langle D \rangle_{B,2}=224\pm9$\,nm, and $\sigma_{B,2}=48\pm7$\,nm. The characteristics of the small nanobubbles are then $D_{B,1}^*=58\pm4$\,nm, $\langle D \rangle_{B,1}=64\pm4$\,nm and $\sigma_{B,1}=29\pm4$\,nm.

The observation of two co-existing but clearly separated sets of bubbles has not been reported or predicted before. The larger nanobubbles are created during the temporal gas oversaturation in the water during the exchange process, in agreement with previous observations~\cite{zhang06a}, while we hypothesize that the smaller ones are formed once the saturated conditions are restored. Notice that the smaller set of bubbles in case B is fairly similar to the population in case A in both the shape of $P(D)$, as well as the order of magnitude of the respective maxima ($40\pm2$\,nm and $58\pm4$\,nm, resp.). Remarkably, these maxima are close to the experimental result of Simonsen~\emph{et al.}~\cite{simonsen04}, who found a normal distribution of sizes with $D^{*}=66\,\mathrm{nm}$ under identical lab conditions (i.e. gas-equilibrated Milli-Q water put on surfaces with a static contact angle $90^{\mathrm{o}}$).

\begin{figure}
\centering
    \includegraphics[width=40mm]{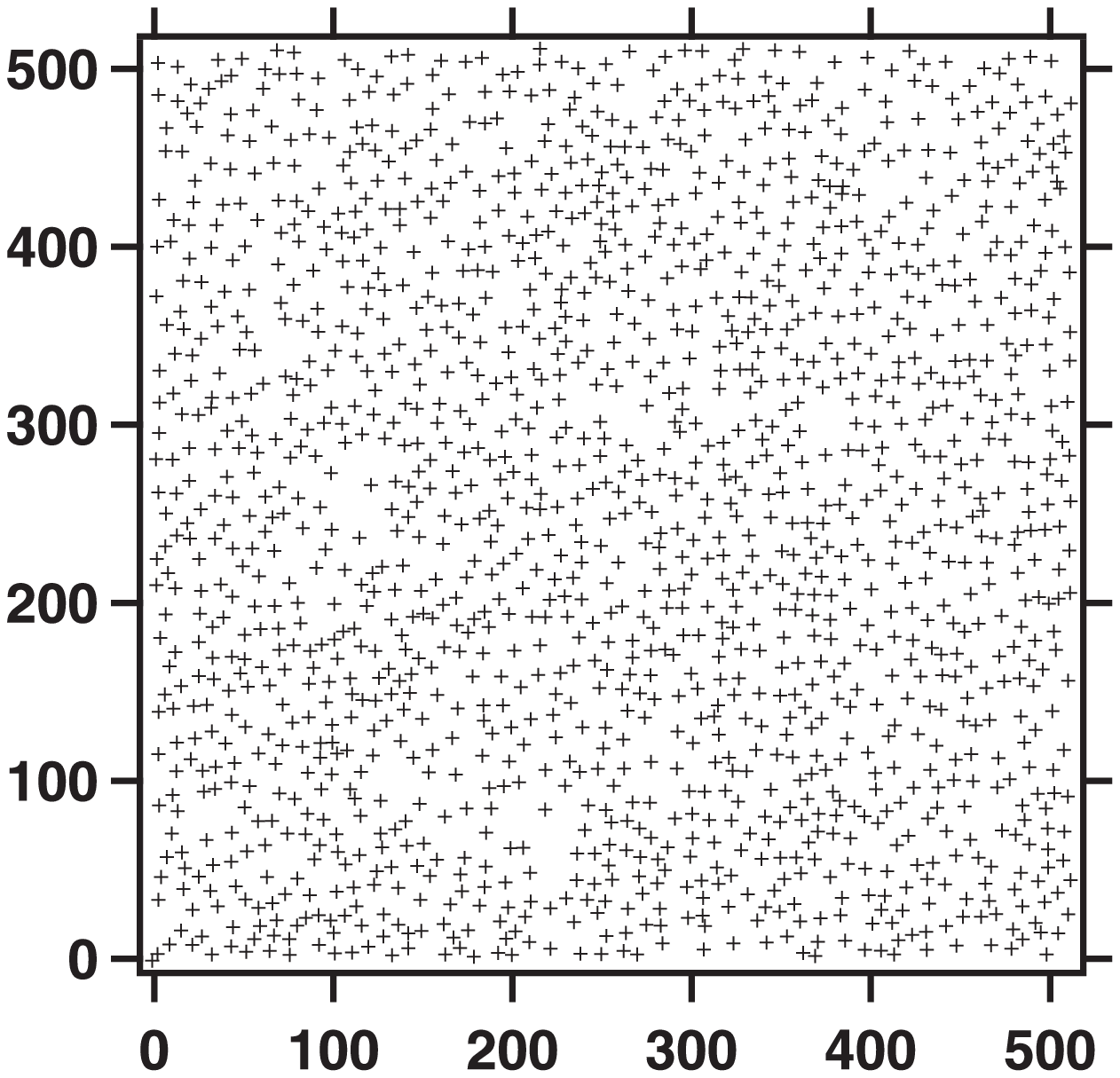}
    \includegraphics[width=40mm]{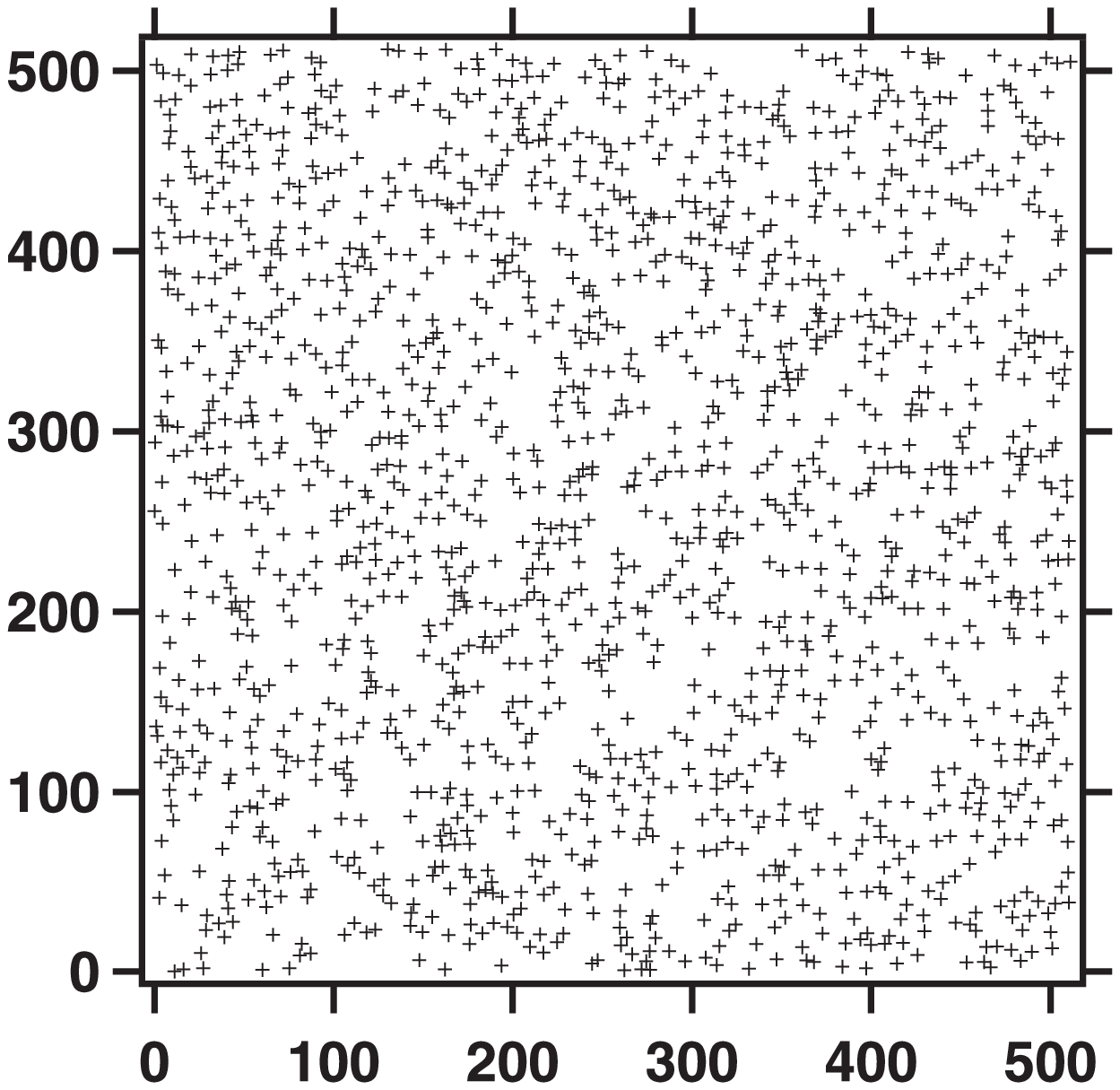}
    \caption{\label{Fig:exp_mc_A2}Positions of the nanobubbles in case A2 (left) vs. Monte-Carlo simulations of a random packing of hard-disks with the same size distribution and density as experiment A2 (right). The experimental positions show much more structure than the simulated bubbles.}
\end{figure}

\begin{figure}
     \includegraphics[width=41mm]{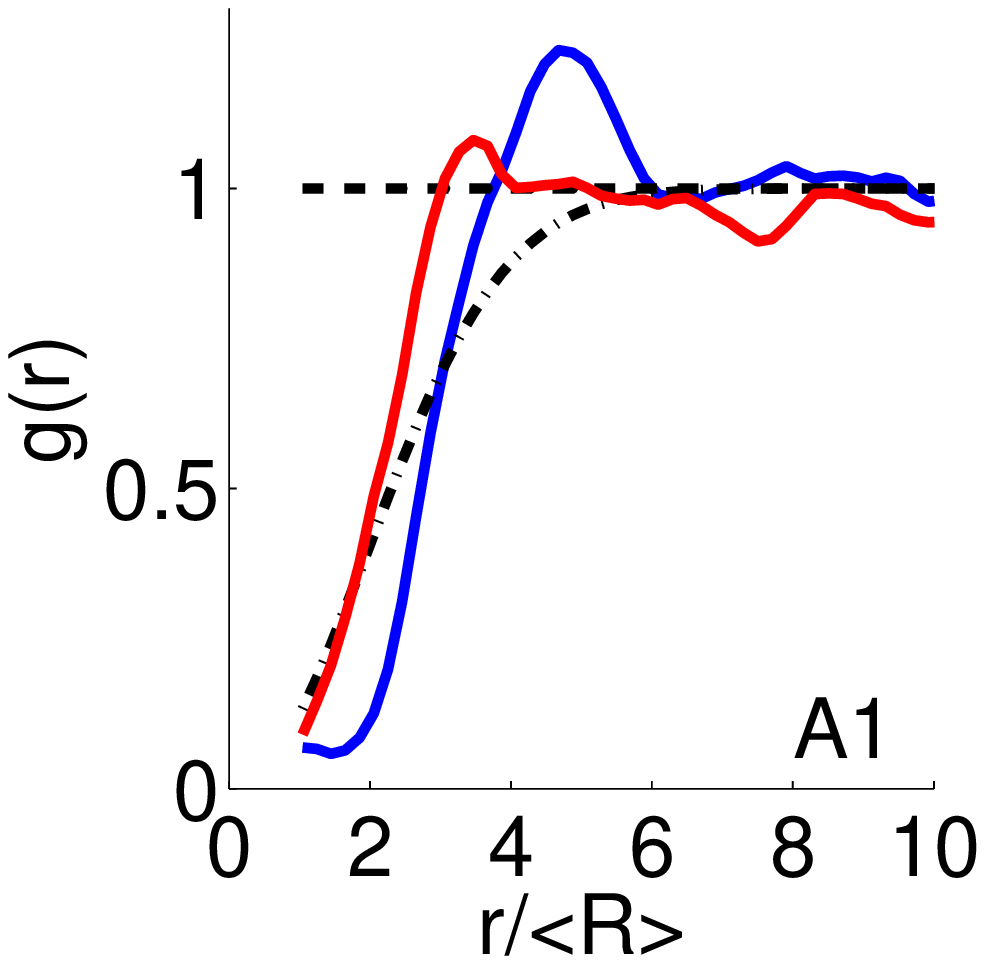}
     \includegraphics[width=41mm]{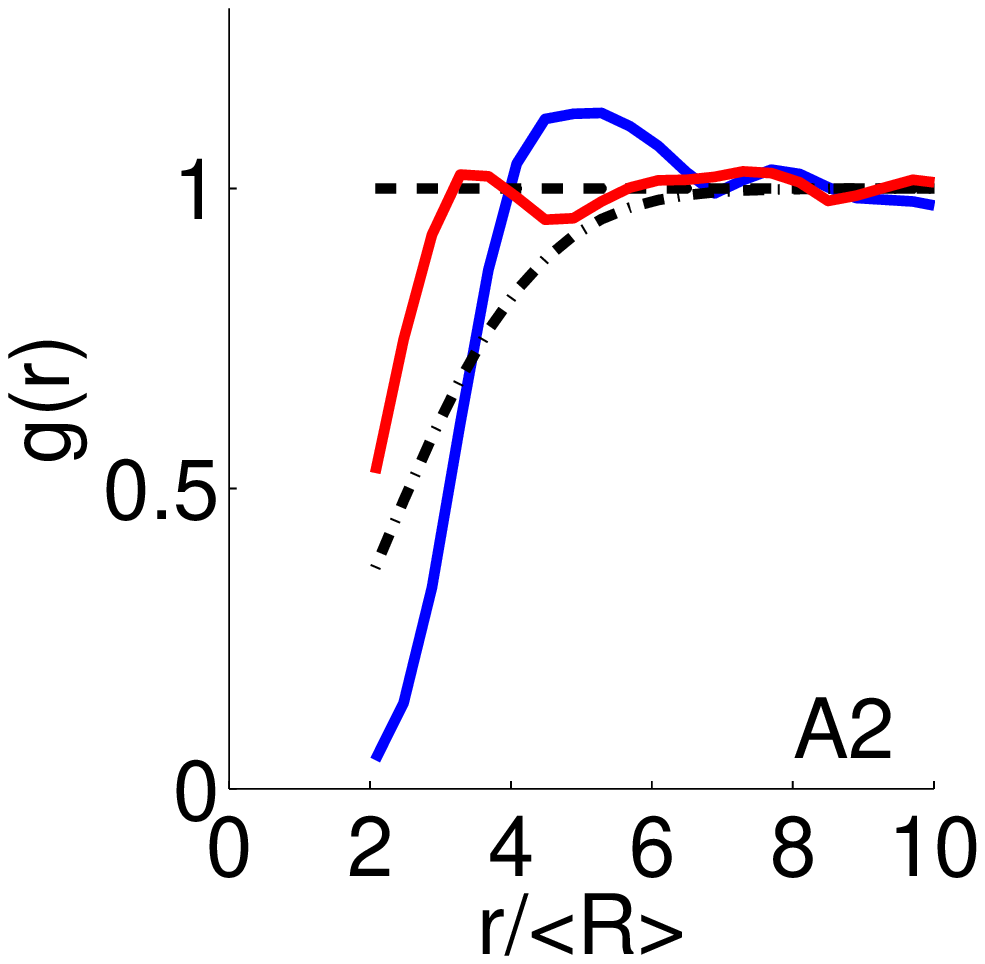}\\
     \includegraphics[width=41mm]{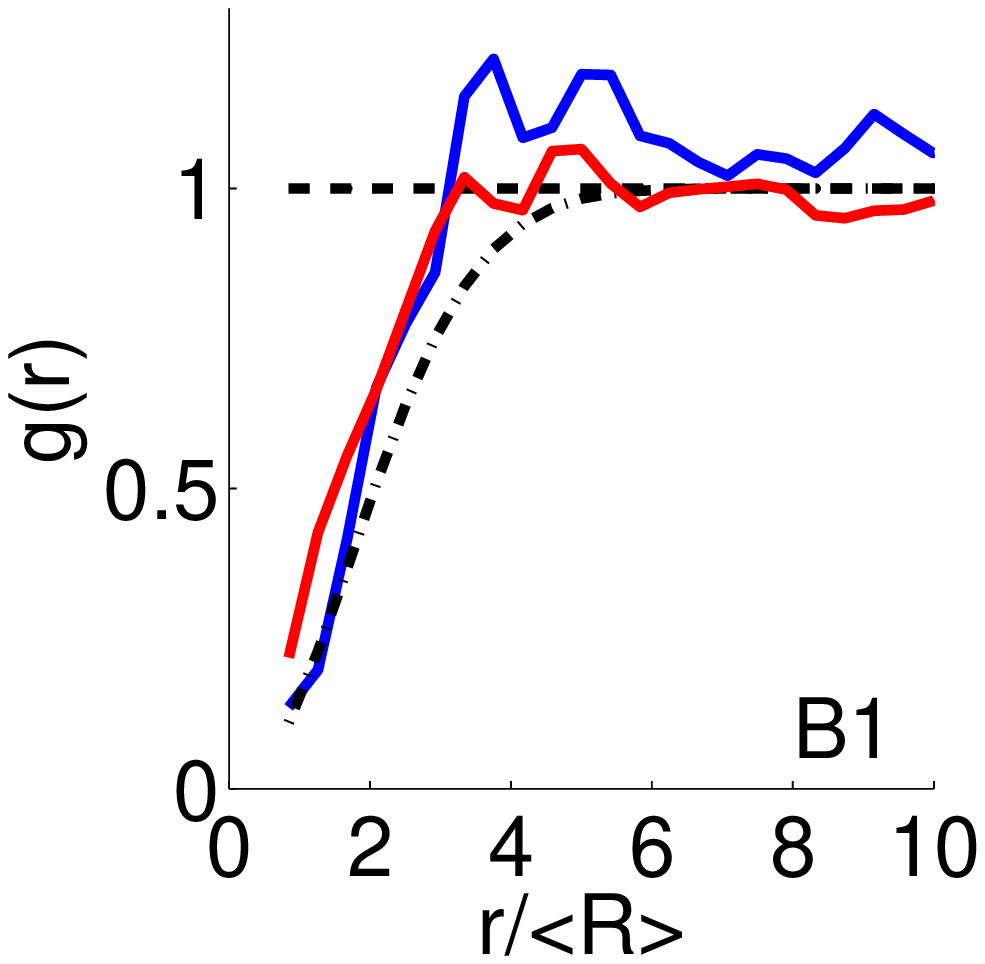}
     \includegraphics[width=41mm]{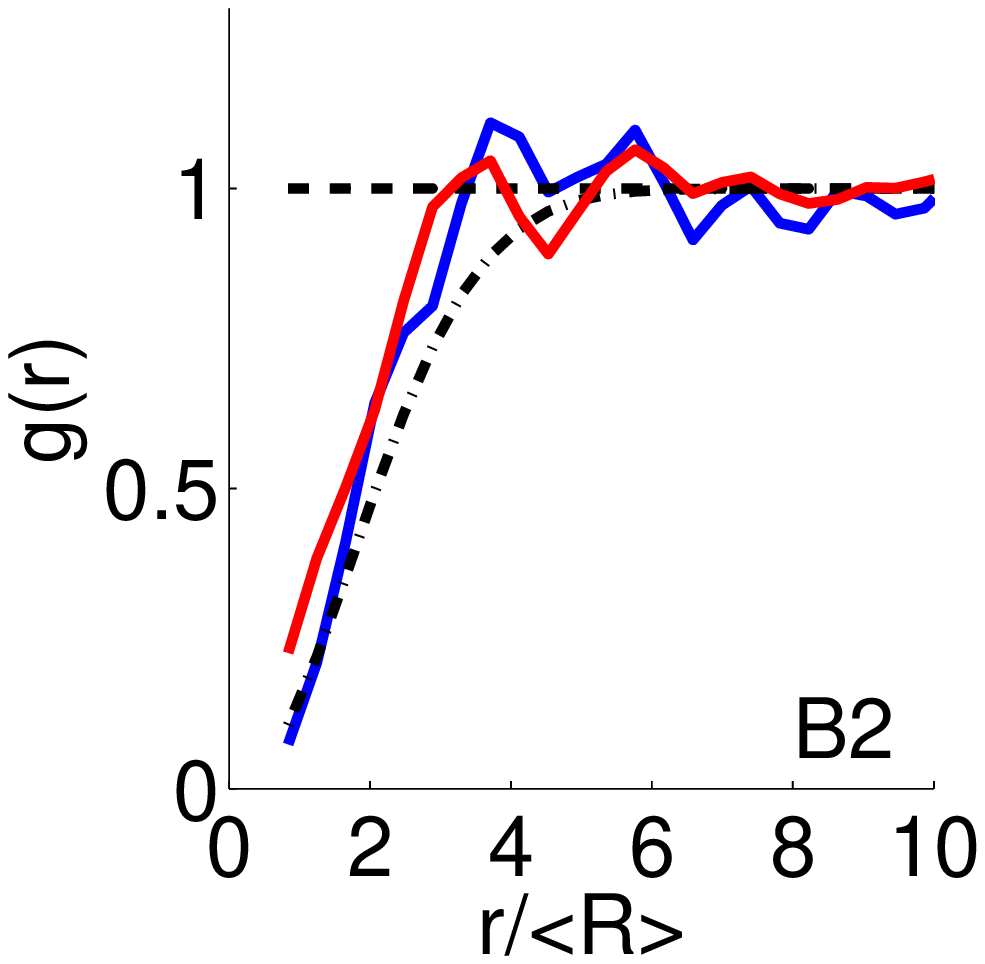}
    \caption{\label{Fig:gr}Radial distribution functions $g(r)$ as a function of $r$ normalized by the mean radius $\langle R \rangle$ for case A (top) and case B (bottom). Blue line: experiment; red line: Monte-Carlo simulations of a random packing of hard-disks with the same size distribution and density as the associated experiments; dashed line: Poisson point process; dash-dotted line: determinantal point process.}
\end{figure}

\begin{figure}
     \includegraphics[width=41mm]{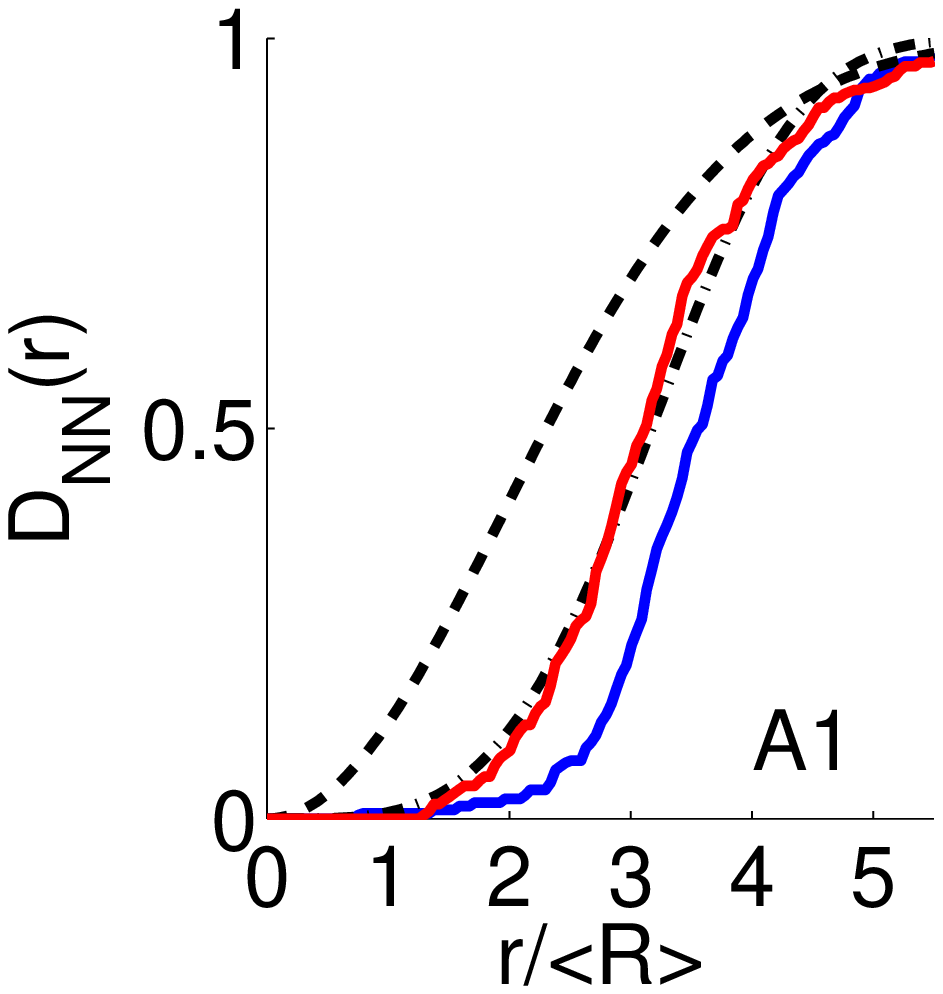}
     \includegraphics[width=41mm]{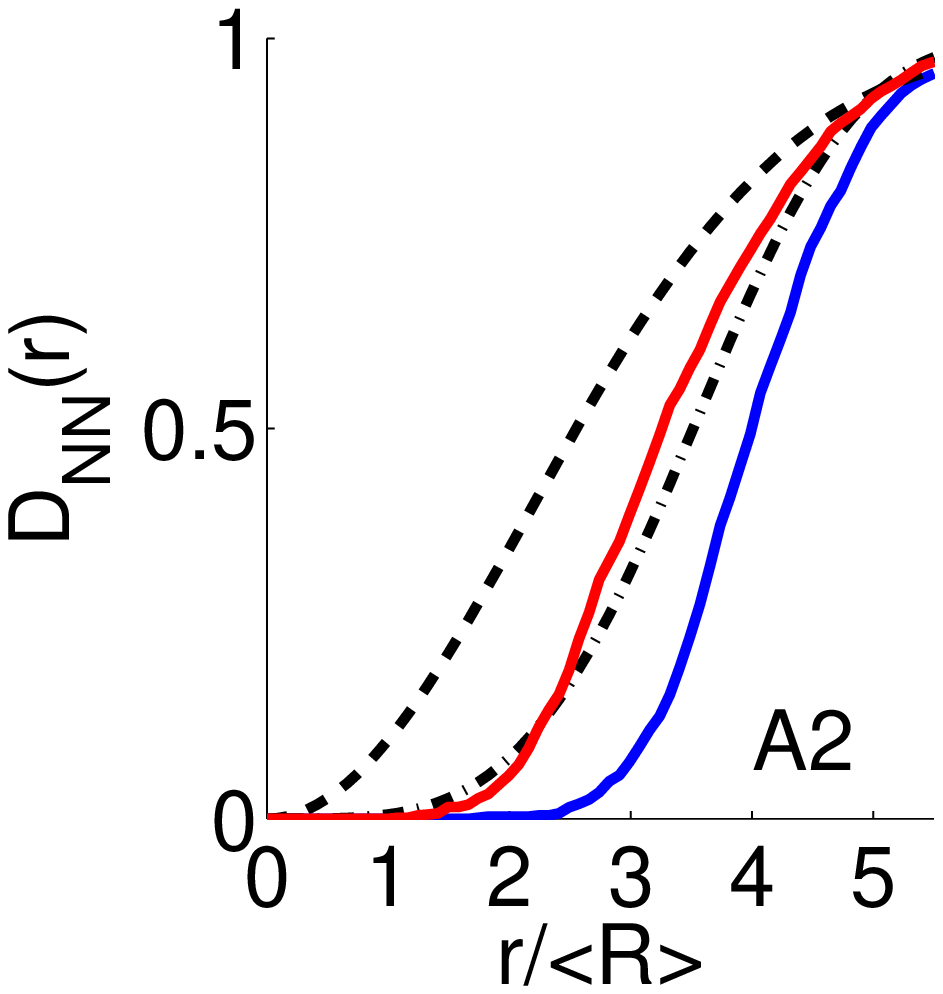}\\
    \includegraphics[width=41mm]{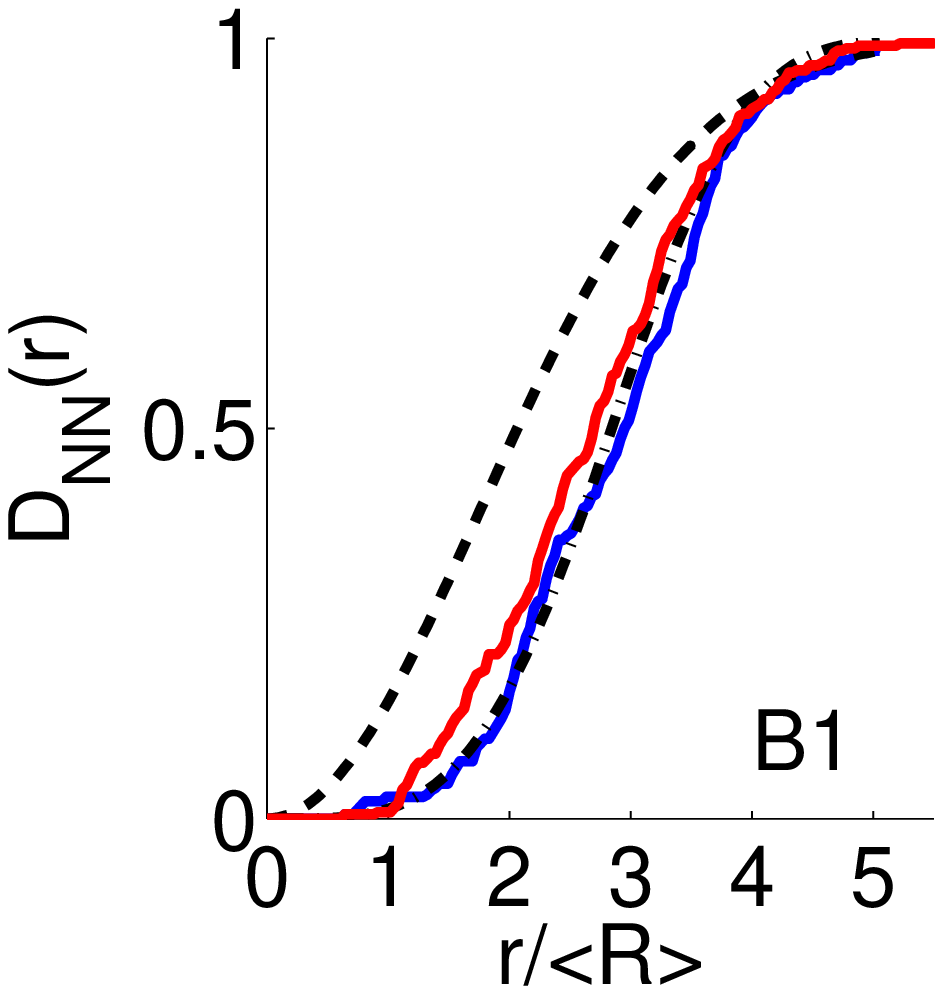}
    \includegraphics[width=41mm]{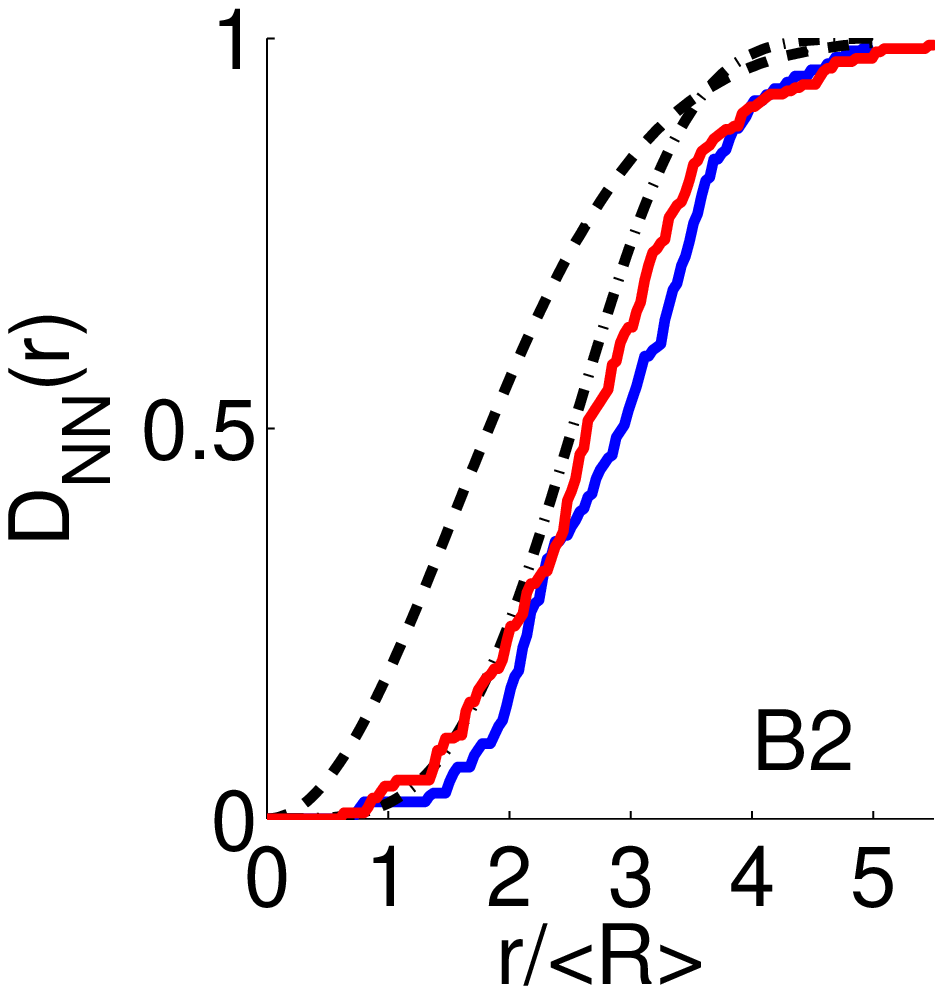}
    \caption{\label{Fig:Dr}Nearest neighbor distributions $D_{NN}(r)$ for the four cases A1--B2. The legend is the same as in Fig.~\ref{Fig:gr}.}
\end{figure}

The homogeneity of the nanobubble coverage depicted in Fig.~\ref{Fig:cases} suggests local structuring of the bubbles. To test this idea quantitatively, Monte-Carlo (MC) simulations of a random packing of hard-disks with the same size distribution and density as in the experiments are employed. For case A2, the nanobubble center positions in both experiment and MC-simulation are depicted in Fig.~\ref{Fig:exp_mc_A2}, which shows that the experimental positions are much more structured than the simulated bubble positions. This effect is further shown by the radial distribution function $g(r)$, which quantifies the probability of finding a bubble at a radial distance $r$ from another bubble, and the nearest neighbor distribution function $D_{NN}(r)$, which gives the probability of finding a nearest neighbor of a nanobubble at a distance less than or equal to $r$~\cite{stoyan95}. The plots of $g(r)$ and $D_{NN}(r)$ are depicted in Fig.~\ref{Fig:gr} and Fig.~\ref{Fig:Dr}, respectively, for both the experimental and MC-simulated positions. In addition, the figures show the distributions for a Poisson point process (where neither steric nor repulsive interaction is present), and a determinantal point process with a very soft repulsion between the points~\footnote{The latter process is a universal 2D point process with explicit expressions of correlation functions of any order (see~\cite{lecaer93,scardicchio09} and references therein). The points are more regularly distributed than they are for a Poisson point process.}.

For case A1 and A2 the experimental curves in Fig.~\ref{Fig:gr} show a significant peak in $g(r)$ at $r \sim 5 \langle R \rangle \sim 5.5 R^*$ while, interestingly, this peak is absent in the corresponding MC-simulations and the determinantal point process. This shows that there is a preferred spacing between the bubbles present in both case A1 and A2, which is not only steric and stronger than the 'soft' repulsion represented by the determinantal point process. The regularity of the bubble positions in case A1 and A2 is also shown in the plots of $D_{NN}(r)$ (Fig.~\ref{Fig:Dr}): the experimental curves are on the right-hand side of the MC-curves. Notice that the $D_{NN}(r)$ curves of the MC simulated cases A1 and A2 (red lines) are close to those given by the determinantal point process (dash-dotted lines) although in the MC simulations hard disks are used without any mutual interaction. The similarity is not seen when the MC simulation utilizes a single disk size. Hence, the effect of the size distribution looks like an effective soft repulsion.

In contrast, for cases B1 and B2 no significant difference is observed between experiments and MC-simulations, in both $g(r)$ and $D_{NN}(r)$. The reason could be that the statistics is too poor: case A counted three times as many bubbles as case B. Another, more likely reason could be that the number densities in case B are too low for structuring effects to be present. In case B1 and B2 the number density was 13.8 per $\mu \mathrm{m}^2$ on average, while in case A1 and A2 this was 70.7 per $\mu \mathrm{m}^2$, more than a factor of five difference.

In summary, it is demonstrated for two types of surface nanobubble populations that nanobubbles (i) show a preference in size, and (ii) show a preference in spacing. The first observation shows up in both cases, while the second observation only shows up when the number densities are large enough. In case A the size distribution is found to be distributed according to a generalized gamma law. A very similar size distribution is present in case B, where in addition a larger set of normal-distributed nanobubbles is present, which were created most likely during the temporal gas-oversaturation in the water. These findings are consistent with the hypothesis of a uniform stabilizing mechanism leading to a preferred radius, as put forward in~\cite{brenner08}. Comparisons with MC simulations show that densely packed nanobubbles do not reside randomly, but choose a position were it is easiest for them to be: away from each others vicinity. The physical mechanism responsible for this effect could be the limited availability of gas in the vicinity of an already formed nanobubble, prohibiting the nucleation of other nanobubbles nearby. Alternatively, nanobubbles could be formed instantaneously from the breakup of a homogeneous gas film into individual bubbles, analogous to the break-up of thin liquid films into surface patterns~\cite{bestehorn01}. Thirdly, the ordering effect could result from a short-range repulsive force, e.g. due to surface charges. Further theoretical and experimental studies are required to unravel the precise formation mechanism of nanobubbles and their mutual interplay at the nanoscale.

We thank H. Zandvliet, S. Kooij, A. Prosperetti and J.H.~Snoeijer for stimulating discussions.
This work was supported by NanoImpuls/NanoNed, the nanotechnology program of the Dutch Ministry of Economic Affairs (Grants TPC.6940, TMM.6413).

\end{document}